\documentclass[aps,twocolumn]{revtex4}
\usepackage{graphicx}
\usepackage{dcolumn}
\usepackage{amsmath}

\begin{document}

\title{Universal Distribution of Centers and Saddles in Two-Dimensional Turbulence}

\author{Michael Rivera}
\affiliation{Department of Physics and Astronomy, University of Pittsburgh,
Pittsburgh, PA 15260}
\author{Xiao-Lun Wu} 
\affiliation{Department of Physics and Astronomy, University of Pittsburgh,
Pittsburgh, PA 15260}
\author{Chuck Yeung}
\affiliation{School of Science, The Pennsylvania State University at Erie, Erie, PA
16563-0203}

\date{\today}

\begin{abstract}
The statistical properties of the local topology of two-dimensional turbulence are investigated using an electromagnetically forced soap film. The local topology of the incompressible 2D flow is characterized by the Jacobian determinant $\Lambda(x,y) ={1 \over 4} (\omega^2 - \sigma^2)$, where $\omega (x,y)$ is the local vorticity and $\sigma (x,y)$ is the local strain rate. For turbulent flows driven by different external force configurations, $P(\Lambda)$ is found to be a universal function when rescaled using the turbulent intensity. A simple model that agrees with the measured functional form of $P(\Lambda)$ is constructed using the assumption that the stream function, $\psi(x,y)$, is a Gaussian random field.
\end{abstract}


\maketitle 

Geometrical and topological ideas have played important roles in our
understanding of turbulent phenomena. For example the energy cascades of both
three-dimensional (3D) and two-dimensional (2D) turbulence are generally
pictured as hierarchical structures of interacting eddies \cite{Frisch}. 
Although topological ideas find wide spread use in the {\em interpretation} of
data and theory, they are not generally used to analyze data
and develop theory. The reason for this is because there currently exist no
measurement technique for the extraction of entire turbulent velocity fields
from a 3D fluid. Since the bulk of turbulence experiments have been performed
in 3D, this lack of enthusiasm for topological approaches is
an artifact of limited data. In 2D, however, the situation is more tractable since particle tracking
can measure entire 2D velocity fields
\cite{Rivera:PRL98}. In this article the statistical properties of the local
topology of 2D velocity fields extracted from a turbulent soap film are
investigated.

\begin{figure}
\includegraphics[width=3.375in]{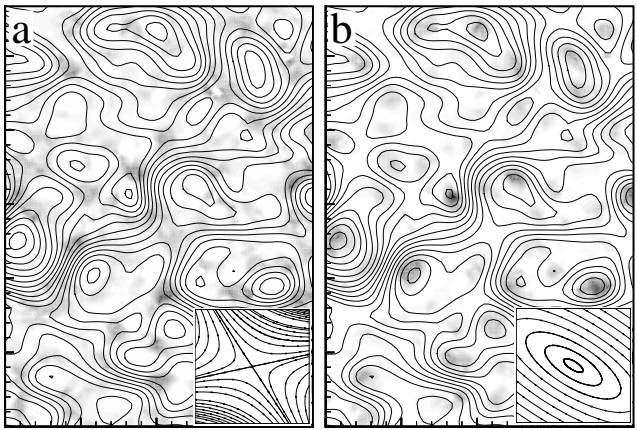}
\caption{$\Lambda$ field in turbulence driven by a linear array of magnets. For clarity, $\Lambda^-$ and $\Lambda^+$ are plotted in (a) and (b) respectively with darker shades indicating larger magnitudes of $\Lambda^{\pm}$. The solid lines are the contours of the stream function, $\psi$, and the large tick marks correspond to distances of $1$ cm. The insets in (a) and (b) show the local flow for a saddle and center, respectively.}
\label{fig: typfields}
\end{figure}

The local flow topology in a 2D fluid can be determined, up to a coordinate
transformation, by two numbers: the Jacobian determinant, $\Lambda(x,y) \equiv
\partial(v_x,v_y)/\partial(x,y)$, and the divergence of the flow field, ${\bf
\nabla} \cdot {\bf v}$. Since the velocity fields under consideration are
essentially incompressible, the single field $\Lambda(x,y)$ uniquely defines
the flow. This quantity lends itself to a simple geometrical interpretation. If
$\Lambda(x,y) < 0$, then in the reference frame of the fluid at point $(x,y)$
the flow structure is similar to that shown in the inset of Fig.\ \ref{fig:
typfields}(a) and is called a saddle. On the other hand, if $\Lambda(x,y)>0$, the
flow structure resembles that in the inset of Fig.\ \ref{fig: typfields}(b) and
is called a center. If $\Lambda(x,y)=0$, the local field is in either a state of
pure shear, or more complicated non-linear shear. The quantity $\Lambda$ also
plays a significant role other than measuring the local topology of a 2D
fluid.  If the 2D fluid has density $\rho$ and experiences external forces that
are divergence free, then $\Lambda$ is proportional to the Laplacian of the
pressure field, $\Delta p= 2\rho \Lambda$.  Therefore saddles are regions of locally high pressure, while centers are regions of
locally low pressure.

One can express $\Lambda$ in terms of two more familiar quantities, the vorticity $\omega$ and the strain rate $\sigma$:
\begin{eqnarray}
\Lambda={1 \over 4}(\omega^2-\sigma^2),
\end{eqnarray}
where $\omega^2=\sum_{i,j}(\partial_iv_j-\partial_j v_i)^2 / 2$ and $\sigma^2=\sum_{i,j}(\partial_iv_j+\partial_j v_i)^2 / 2$. Hence,
regions of large positive $\Lambda$ correspond to strong vorticity, whereas
regions of large negative $\Lambda$ correspond to strong elongation.  Note that
this is not exclusive, meaning that saddle regions could also have a large vorticity
provided the local strain is still larger. Similarly, centers could be strongly
elongated as long as their rotation rate exceeds the strain rate.

This article will investigate the probability $P(\Lambda)$ that a point in turbulent flow is
behaving as a saddle or a center of a given strength.  This will be done in 2D using an
electromagnetically forced soap film (e-m cell) \cite{Rivera:PRL00}. Theoretical predictions of the behavior of $P(\Lambda)$ are also developed in this article and shown to be in
agreement with most of the features observed in the measurement.

A description of the operation of the
e-m cell can be found in \cite{Rivera:PRL00} and will not be reviewed in depth
here. Briefly, the e-m cell creates a state of energetically steady turbulence
in a  $50$ $\mu$m thick soap film using electromagnetic fields. Data
from the e-m cell has been shown to be consistent with exact predictions that
can be made using the forced 2D incompressible Navier-Stokes equation:
\begin{eqnarray}
	\frac{\partial {\bf v}}{\partial t} + {\bf v}
	\cdot {\bf v} \mbox{\boldmath{$\nabla$}} {\bf v}
	= -\mbox{\boldmath{$\nabla$}} p + \nu \nabla^2 {\bf v} + {\bf F} - \alpha 	{\bf v},
\label{eq: Navier-Stokes}
\end{eqnarray}
Here, ${\bf v}$ is the velocity field, $p$ is the internal pressure field
normalized by the fluid density, and ${\bf F}$ is the electromagnetic force
field acting on the fluid. The constant $\nu\simeq 0.016$ cm$^2$/s is the
kinematic viscosity. The last term in Eq.\ \ref{eq: Navier-Stokes}, $-\alpha
{\bf v}$, accounts for the frictional effect of air acting on the soap film. This system has two dimensionless control parameters which can be
varied by changing either the magnitude of ${\bf F}$ or $\alpha$. The first is
the Reynolds number defined as $Re = v_{rms}/k_{inj}\nu$ and the second is the
dimensionless air-friction $\gamma = \alpha/k_{inj}^2 \nu$. Here $k_{inj}$
is a typical wavenumber describing the spatial variation of the electromagnetic
force field, called the injection wavenumber. In these
experiments, $\gamma$ assumes values from 0.1 to 1, corresponding to a regime
of weak to moderate damping, and $Re$ varies from 10 to 100. 

To establish that the local topology measurements are not merely a reflection
of the geometrical aspects of the external forcing two
different force configurations were used. Both force fields were
unidirectional with $F_y=0$. The $F_x$ for each field varied in different
spatial patterns which will be called linear and square. $F_x$ is given
by $F_x  =  F_0\sin(k_0y)$ for the linear field, and $F_x  = 
F_0\sin(k_0y/\sqrt{2})~ \sin(k_0x/\sqrt{2})$ for the square field. For both cases, $k_{inj}=k_0=2\pi/a$, with $a \simeq 0.6$ cm for the linear forcing and $a \simeq 0.8$ cm for the square forcing. Unless otherwise noted the magnitude of the force fields, $F_0$, varied in time with a square waveform of $3$ Hz. Finally,
topological aspects of decaying turbulence were also investigated by driving
the cell and suddenly switching off the drive. It is generally believed that
decaying turbulence is very different from forced turbulence due to the absence of an
inverse energy-cascade range \cite{Chasnov:PFL97}.

The velocity field is measured by particle tracking velocimetry \cite{Rivera:PRL00}, which
typically yields 5000 velocity vectors per image. Figure \ref{fig: typfields}
is a typical $\Lambda$ field calculated from a single velocity field driven
by the linear forcing with $v_{rms} \simeq 10$ cm/s. For clarity, the
distribution of centers $\Lambda>0$, denoted as $\Lambda^+$, is plotted
separately from the distribution of saddles $\Lambda<0$, denoted as
$\Lambda^-$. Unlike the velocity fields, which tend to be spatially extended, the distribution
of $\Lambda$ appears to be localized or spotty, more so for centers than for
saddles.

Sequences of 400-500 $\Lambda$ fields were acquired for various values of
$\gamma$ and $Re$, and for different types of forcing (see Table \ref{tab: constants}). From these,
probability distribution functions (PDF) of $\Lambda$ were computed, and the results are shown in Fig.\ \ref{fig:
lambdadist}. All of the PDFs have a strikingly similar form. One observes that
the PDF is asymmetric with a nearly exponential decay for $\Lambda<0$ and an
initially quick, perhaps stretched exponential, decay for small $\Lambda>0$,
eventually relaxing to what appears to be exponential decay for large $\Lambda>0$.
The anomalously long tail of the PDF for $\Lambda^+$ is consistent with the
visual appearance of velocity fields in that powerful vortices are more
conspicuous than saddles in the flow.   Finally, we note that the PDF is
non-analytic at $\Lambda=0$.

\begin{table}
\begin{tabular}{||c|c|c|c|c|c|c|c||}\hline\hline
forcing type & $Re$ & $\gamma$ & $\tau$ (s) & $b$ & $\tilde{\Lambda} (s^{-2})$ & $\zeta$ &symbol\\ \hline
linear & 60 & 0.28 & na & 0.31 &712&2.53&$\diamond$\\
linear & 60 & 0.56 & na & 0.31 &1014&2.60&$\triangleright $\\
linear & 60 & 0.97 & na & 0.31 &1282&2.54&$\circ $\\
square & 80 & 0.6 & na & 0.4 &898&2.84&$\star $\\
decay &70 &0.5 & 0.5 & 0.35 &104&2.54&$\triangleleft $\\
decay &70 &0.5 & 1.0 & 0.35 &39.1&3.4&$\Delta $\\ \hline \hline
\end{tabular}
\caption{Global constants for different types of forcing in the e-m cell.   Here, $\tau$ is the amount of time the system was allowed to decay before taking data. The symbols correspond to those plotted in the later figures.  $Re$ and $\gamma$ values listed for decaying turbulence denote the initial values of these parameters.}
\label{tab: constants}
\end{table}

\begin{figure}
\includegraphics[width=3.375in]{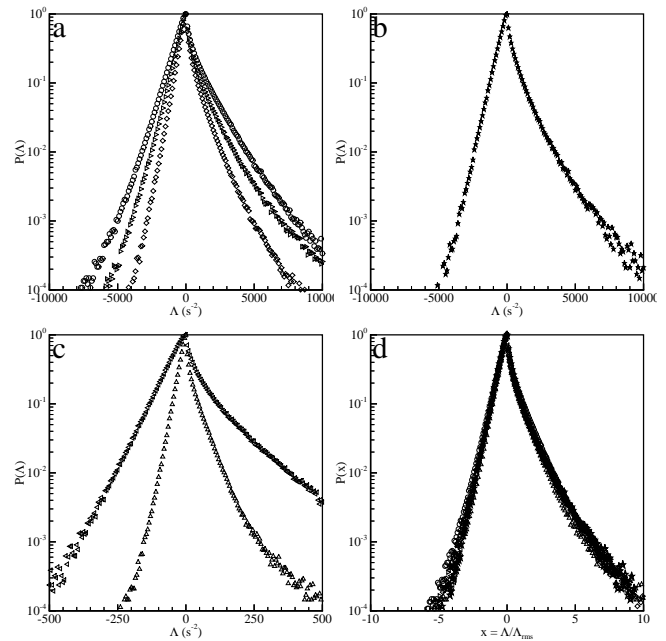}
\caption{(a) The distribution function $P(\Lambda)$ for: (a) linearly forcing with $Re \approx 60$ and $\gamma$ values  of $0.28$ ($\diamond $), $0.56$ ($\triangleright $) and $0.97$ ($\circ $), (b) square forcing with $Re \approx 80$ and $\gamma \approx 0.6$, and (c) decaying turbulence at two separate times, $0.5$ s ($\triangleleft$) and $1.0$ s ($\Delta$) after the forcing is turned off. All of these data sets are plotted together in (d) after normalizing by the respective $\Lambda_{rms}$.}
\label{fig: lambdadist}
\end{figure}

The PDFs in Fig.\ \ref{fig: lambdadist}(a)-(c) can be collapsed on a single
master curve if $\Lambda$ is normalized by its RMS value, $x=\Lambda/
\Lambda_{rms}$.
As seen in Fig.\ \ref{fig: lambdadist}(d), over four decades in the vertical
scale, the quality of the data collapse is excellent. The agreement between decaying and driven
turbulence also indicates that $P(\Lambda)$ is independent of the presence or
lack of an inverse cascade. These observations lead us to believe that
the functional form of $P(\Lambda)$ is universal for 2D turbulence, though the flow appears to
be visually different.  

The asymmetry of $P(\Lambda)$ is clearly reflected in the moments of the PDF. 
All measurements show that $\langle \Lambda \rangle\approx0$, yielding $\langle \Lambda^- \rangle \approx - \langle \Lambda^+\rangle$.  However, the second moment is tilted toward the positive
side with $\zeta \equiv \langle (\Lambda^+)^2\rangle / \langle (\Lambda^-)^2\rangle \sim 2.5$ (see Table \ref{tab: constants}). Though the data is not presented here, it should be noted that these results in the turbulent regime are markedly different from those obtained in the laminar flow regime
with an ordered array of vortices. In this case, the PDF is nearly symmetric
with $\langle \Lambda^- \rangle = - \langle \Lambda^+ \rangle$ and $\langle
(\Lambda^+)^2 \rangle \approx \langle (\Lambda^-)^2 \rangle$.

\begin{figure}
\includegraphics[width=3.in]{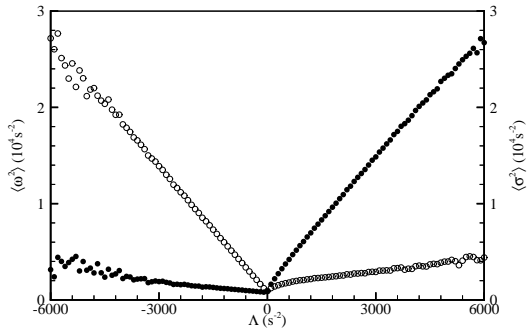}
\caption{The conditional expectation values of  $\sigma^2$ ($\circ$) and
$\omega^2$ ($\bullet$) for a given $\Lambda$.  The measurement is taken
from a linearly forced flow at $Re = 60$ and $\gamma = 0.56$.}
\label{fig: conditionalprobs}
\end{figure}

We also considered the conditional probability of enstrophy $\omega^2$ and
square strain rate $\sigma^2$ for a given $\Lambda$. We find that the conditional
PDF for both $\omega^2$ and $\sigma^2$ are
approximately exponential. The conditional expectation values for $\langle
\sigma^2 \rangle$ and $\langle \omega^2 \rangle$ for a given $\Lambda$ are
plotted in Fig.\ \ref{fig: conditionalprobs}. It is noticeable that while $\langle
\sigma^2 \rangle$ and $\langle \omega^2 \rangle$ depend linearly on $\Lambda$ for $\Lambda<0$, such a linear relation does not hold for small positive $\Lambda$. 

The universal nature of $P(\Lambda)$ indicates that the probability distribution can be understood from basic principles and minimal models without requiring a detailed model of the flow. Since the experiment approximates 2D
incompressible flow, the velocity can be written in terms of a stream function $\psi(x,y)$ with $v_x = \partial_y \psi$ and $v_y = -\partial_x \psi$. The Jacobian determinant is then $\Lambda = - (\partial_x \partial_y
\psi)^2 + ( \partial_x^2 \psi ) ( \partial_y^2 \psi )$.  

We can show that $\langle \Lambda \rangle = 0$ if the correlation function of $\psi$ is translationally invariant, that is, $\langle \psi(x,y) ~ \psi(x',y') \rangle = M(u,w)$ is a function of $u = |x-x'|$ and $w= |y - y'|$ only. If we further assume that $M(u,w)$ is analytic at $u = w = 0$, we find
\begin{eqnarray}
\langle \Lambda \rangle &=& 	\langle -(\partial_x v_x) ( \partial_y v_y) 	+ (\partial_x v_y) ( \partial_y v_x )\rangle\nonumber \\ 	&=&-\langle \partial_{u} \partial_{w}\partial_{w} \partial_{u} M \rangle + \langle \partial_{u}^2 \partial_{w}^2 M \rangle = 0.
\end{eqnarray}
Note that all derivatives of $M$ here and below are evaluated at $u=w=0$.  We expect that $\langle \Lambda \rangle = 0$ for a wide variety of flows since we do not assume that $M$ is isotropic, simply that it is translationally invariant.

To understand the PDF in the turbulent regime,  we need a further assumption. The simplest assumption  is that the stream function is a Gaussian random field. The derivatives of a Gaussian field are also Gaussian so that $\partial_x v_x$, $\partial_y v_y$, $\partial_y v_x$ and $\partial_x v_y$ are all Gaussian random fields with their cumulants given by 4th order derivatives of $M(u,w)$. Hence it is straightforward to calculate $P(\Lambda)$, since $\Lambda$ is a functional of Gaussian fields.

The PDF, $P(\Lambda)$, is found to depend on only two positive parameters. The first parameter, $\tilde{\Lambda}=\sqrt{ \langle (\partial_x v_y)^2 \rangle \langle (\partial_y v_x)^2 \rangle} = \sqrt{ (\partial_u^4 M ) ~ ( \partial_w^4 M) }$ simply sets the scale of $\Lambda$ and hence does not affect the shape of the distribution.  The second parameter $b$ is
\begin{equation}
b=-\frac{\langle 	(\partial_x v_y) (\partial_y v_x) \rangle }{ 	\sqrt{ \langle (\partial_x v_y)^2 \rangle ~ 	\langle (\partial_y v_x)^2 \rangle}}= 	\frac{ \partial_u^2 \partial_w^2 M }{ 	\sqrt{ (\partial_u^4 M) ~ 	( \partial_w^4 M ) }}.
\end{equation}
Hence $b$ is the normalized correlation of the shear strain rates in the different directions and is the only parameter that affects the shape of $P(\Lambda)$. In principle $b$ is a free parameter.  However, if $M(u,w)$  has elliptical symmetry, that is $M(u,w) = M( u^2 + c w^2)$ where $c$ is a constant, $b$ assumes the value $1/3$.  

Introducing a rescaled $\Lambda$ via $\Lambda' =\Lambda/\tilde{\Lambda}$, the
PDF can be written in terms of an integral \cite{Chuck:01}:
\begin{widetext}
\begin{eqnarray}
	P(\Lambda') = 
		C\exp\left(\frac{-\Lambda'}{1+b}\right)
		\int_{r_{min}}^{\infty}dr
		\int_0^{2\pi} d\phi ~ {r \over  \sqrt{\Lambda'+r^2}}
		\exp\left[-r^2 \left(\frac{1}{1+b} +
		\frac{\sin^2\phi }{1-b}  +
		\frac{\cos^2\phi}{2b} \right)\right],
\end{eqnarray}
\end{widetext}
where $r_{min}=0$ if $\Lambda'>0$ and $r_{min}=\sqrt{-\Lambda'}$ if $\Lambda'
<0$, and C is a normalization constant. 
For general $b$, this integral must be done numerically.  However,
for $b = 1/3$ we obtain a close form:
\begin{eqnarray}
	P( \Lambda' ) & = &
	\left\{ \begin{array}{ll}
		\frac{2 C \pi^{3/2}}{3}  
		e^{3 \Lambda'/2} 	
				& \mbox{for} ~ \Lambda' < 0,
				 \\
		\frac{2 C \pi^{3/2}}{3}
		e^{3 \Lambda'/2} \Phi_{c}\left( \frac{3}{2} \sqrt{ \Lambda' } 
			\right) & \mbox{for}~\Lambda' > 0,
				\end{array}
				\right. 
\label{eq: pdf}
\end{eqnarray}
where $\Phi_{c}$ is the complemetary error function. The different lower 
limits of integration, $r_{min}$, for positive and negative $\Lambda'$ makes
$P(\Lambda)$ asymmetric and non-analytic at zero. For negative $\Lambda'$,
$P(\Lambda')$ is exponential while for positive $\Lambda'$, $P(\Lambda')$ has
a non-exponential behavior for small $\Lambda'$ followed by an exponential decay at
large $\Lambda'$.  The asymptotic exponential decay rates $k_{\pm}$ can be
obtained for general $b$ with $k_{+} = 1/(1+b)$ and $k_{-}= 1/(2b)$.  Since $b < 1$, $k_{\_} > k_+$ and
the asymptotic decay is faster for negative $\Lambda'$.

\begin{figure}
\includegraphics[width=3.in]{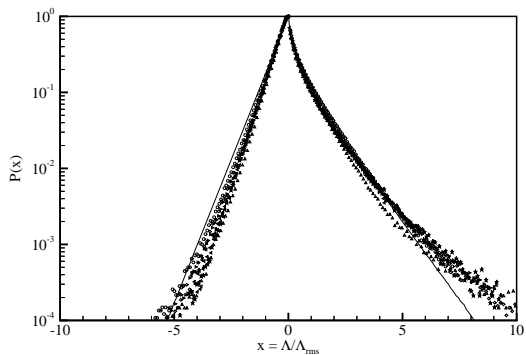}
\caption{The same rescaled PDF as in Fig. 2(d) together with the theoretical 
PDF calculated from Eq. \ref{eq: pdf} with $b=1/3$ (solid line).}
\label{fig: theoryfit}
\end{figure}

To test this model, $b$ was measured directly from the experimental flow fields and found to be close to the expected value of $1/3$; ranging between a low value of $0.3$ for linear forcing to a high value of $0.4$ for the square forcing (see Table \ref{tab: constants}).
The scaled PDF determined from Eq. \ref{eq: pdf} for $b = 1/3$
together with the experimental PDFs are displayed in Fig.\ \ref{fig: theoryfit}. 
The assumption that the stream function is a
Gaussian field captures all the main features of
the experimental PDF.  
In particular, the theory reproduces the asymmetry between
positive and negative $\Lambda$, and
the exponential form of $P(\Lambda)$ for negative $\Lambda$.
For $\Lambda > 0$, the theory predicts the pronounced curvature on
the semi-log plot in agreement with the experiment.  For larger
$\Lambda$ the theory predicts that the PDf decays more slowly
for positive than for negative $\Lambda$ in agreement with observations.
However, the experiment shows a much more pronounced curvature in the
large positive $\Lambda$ tail than the Gaussian model.
This may indicate a breakdown of the Gaussian assumption in regions of large vorticity.

To summarize, the PDF of the Jacobian determinant in 2D turbulence deviates
strongly from Gaussian behavior with prominent exponential tails for large
$|\Lambda|$. Moreover, $P(\Lambda)$ is non-analytic at $\Lambda = 0$ and decays
more slowly for positive $\Lambda$ or the centers. This is evidenced by the
ratio $\langle (\Lambda^+) ^2 \rangle/\langle (\Lambda^-) ^2 \rangle
\sim 2.5$. The non-analytic nature stems from the fact that centers and
saddles are topologically distinct and cannot transform continuously from one
to the other.  The experimental findings reported here are universal,
independent of turbulent intensity and the means of turbulence generation. That
this behavior is universal may be linked to the fact that $\Lambda$ is a local
quantity and therefore insensitive to the long-range spatial correlations of
the turbulent velocity field. The locality of $\Lambda$ may also explain the
success of the Gaussian approximation for the stream function $\psi(x,y)$. This
Gaussian  assumption may prove useful in investigating other features of 2D
turbulence including 2D scalar turbulence.

M. Rivera and X.L. Wu acknowledge support from NASA. 
C.~Yeung acknowledge support from NSF Grant DMR-9986879 and 
Research Corporation Grant CC3993. We also would like to thank D. Jasnow, J.
Vi\~nals and W.I. Goldburg for insightful discussions and Kim and
Karen Herrmann for careful reading of the manuscript.

\bibliography{topology}

\begin{thebibliography}{5}
\expandafter\ifx\csname natexlab\endcsname\relax\def\natexlab#1{#1}\fi
\expandafter\ifx\csname bibnamefont\endcsname\relax
  \def\bibnamefont#1{#1}\fi
\expandafter\ifx\csname bibfnamefont\endcsname\relax
  \def\bibfnamefont#1{#1}\fi
\expandafter\ifx\csname citenamefont\endcsname\relax
  \def\citenamefont#1{#1}\fi
\expandafter\ifx\csname url\endcsname\relax
  \def\url#1{\texttt{#1}}\fi
\expandafter\ifx\csname urlprefix\endcsname\relax\def\urlprefix{URL }\fi
\providecommand{\bibinfo}[2]{#2}
\providecommand{\eprint}[2][]{\url{#2}}

\bibitem[{\citenamefont{Frisch}(1995)}]{Frisch}
\bibinfo{author}{\bibfnamefont{U.}~\bibnamefont{Frisch}},
  \emph{\bibinfo{title}{Turbulence: The Legacy of A. N. Kolmogorov}}
  (\bibinfo{publisher}{Cambridge University Press, London},
  \bibinfo{year}{1995}).

\bibitem[{\citenamefont{Rivera et~al.}(1998)\citenamefont{Rivera, Vorobieff,
  and Ecke}}]{Rivera:PRL98}
\bibinfo{author}{\bibfnamefont{M.}~\bibnamefont{Rivera}},
  \bibinfo{author}{\bibfnamefont{P.}~\bibnamefont{Vorobieff}},
  \bibnamefont{and} \bibinfo{author}{\bibfnamefont{R.}~\bibnamefont{Ecke}},
  \bibinfo{journal}{Phys. Rev. Lett.} \textbf{\bibinfo{volume}{81}},
  \bibinfo{pages}{1417} (\bibinfo{year}{1998}).

\bibitem[{\citenamefont{Rivera and Wu}(2000)}]{Rivera:PRL00}
\bibinfo{author}{\bibfnamefont{M.}~\bibnamefont{Rivera}} \bibnamefont{and}
  \bibinfo{author}{\bibfnamefont{X.~L.} \bibnamefont{Wu}},
  \bibinfo{journal}{Phys. Rev. Lett.} \textbf{\bibinfo{volume}{85}},
  \bibinfo{pages}{976} (\bibinfo{year}{2000}).

\bibitem[{\citenamefont{Chasnov}(1997)}]{Chasnov:PFL97}
\bibinfo{author}{\bibfnamefont{J.~R.} \bibnamefont{Chasnov}},
  \bibinfo{journal}{Phys. Fluids} \textbf{\bibinfo{volume}{9}},
  \bibinfo{pages}{171} (\bibinfo{year}{1997}).

\bibitem[{\citenamefont{Yeung}(2000)}]{Chuck:01}
\bibinfo{author}{\bibfnamefont{C.}~\bibnamefont{Yeung}}, \bibinfo{journal}{In
  preparation}  (\bibinfo{year}{2000}).

\end{thebibliography}

\end{document}